\documentclass{article}
\usepackage{epsf}
\usepackage{emulateapj,pstricks}
\input{psfig}
\usepackage{mymathptm}
\usepackage{flushrt}

\lefthead{Harris et al.}

\righthead{CHANDRA X-RAY DETECTION OF THE RADIO HOTSPOTS OF 3C295}

\slugcomment{submitted to {\em The Astrophysical Journal}}
\def\ref#1{\noindent\hangindent=24.0pt\hangafter=1{#1}\par}
\def\la{\hbox{\rlap{$<$}\lower.5ex\hbox{$\sim$}\ }}
\def\ga{\hbox{\rlap{$>$}\lower.5ex\hbox{$\sim$}\ }}

\begin{document}

\lefthead{Harris et al.}
\righthead{Chandra X-ray Detection of the Radio Hotspots of 3C295}

\slugcomment{submitted to {\em The Astrophysical Journal}}

\title{
CHANDRA X-RAY DETECTION OF THE RADIO HOTSPOTS OF 3C295}

\author{D. E. Harris,\altaffilmark{1} P.~E.~J.~Nulsen,\altaffilmark{1,2}
T.~J.~Ponman, \altaffilmark{1,3} M. Bautz,\altaffilmark{4}
R. A. Cameron,\altaffilmark{1} L.~P.~David,\altaffilmark{1}
R.~H.~Donnelly,\altaffilmark{1} W.~R.~Forman,\altaffilmark{1}
L.~Grego,\altaffilmark{1} M.~J.~Hardcastle,\altaffilmark{5}
J.~P.~Henry,\altaffilmark{6} C.~Jones,\altaffilmark{1}
J.~P.~Leahy,\altaffilmark{7}, M.~Markevitch,\altaffilmark{1}
A.~R.~Martel,\altaffilmark{8} B.~R.~McNamara,\altaffilmark{1}
P.~Mazzotta,\altaffilmark{1} W.~Tucker, \altaffilmark{1}
S.~N.~Virani,\altaffilmark{1} and J.~Vrtilek\altaffilmark{1}}

\bigskip

\doublespace

\begin{abstract}

An observation of the radio galaxy 3C295 during the calibration phase
of the Chandra X-ray Observatory reveals X-ray emission from the core
of the galaxy, from each of the two prominent radio hotspots, and from
the previously known cluster gas.  We discuss the possible emission
processes for the hotspots and argue that a synchrotron self-Compton
model is preferred for most or all of the observed X-ray emission.
SSC models with near equipartition fields thus explain the X-ray
emission from the hotspots in the two highest surface brightness FRII
radio galaxies, Cygnus A and 3C295.  This lends weight to the
assumption of equipartition and suggests that relativistic protons do
not dominate the particle energy density.

\end{abstract}

{\it Subject headings:}  galaxies: individual (3C295) -- magnetic fields -- radiation mechanisms: non-thermal
\vfill

\section{Introduction}

X-ray emission from knots and hotspots in radio jets has been detected
in only a handful of objects.  The three processes normally considered
for X-ray emission from these features are synchrotron, thermal, and
synchrotron self-Compton (SSC) emissions.  The hotspots of 3C295
(Taylor \& Perley 1992) have radio brightness temperatures comparable
to those of Cygnus A but are so close to the nucleus that previous
X-ray systems could not resolve them.  However, the Chandra X-ray
Observatory$^9$ has the ability not only to separate the
emission from the 3C295 cluster gas, core, and hotspots, but in
addition, it allows us to obtain spectra of each component.

SSC models predict X-ray intensities in agreement with those observed
only for the case of the hotspots of Cygnus A (Harris, Carilli, \&
Perley 1994).  The fact that the
observed X-ray flux agrees with the calculated SSC flux for a magnetic
field strength equal to the classical estimate from equipartition
lends credence to the SSC model but does not prove it.  For all the
other previously detected knots and hotspots, the predicted SSC flux
falls well short of the observed flux, often by two or three orders of
magnitude.

In this paper we present Chandra observations of 3C295, describe the
basic results, and evaluate the emission process for the X-rays from
the radio hotspots. We use H$_0$=70 km s$^{-1}$ Mpc$^{-1}$,
$\Omega_{\Lambda}$=0.7, and $\Omega_{M}$=0.3.  At a
redshift of 0.461, the luminosity distance is 2564 Mpc and
1$^{\prime\prime}$=5.8 kpc.  In discussing 

\smallskip

\footnotesize

$^1$Harvard-Smithsonian Center for Astrophysics, 60 Garden St. Cambridge, MA 02138

$^2$Department of Engineering Physics, University of Wollongong, Wollongong NSW 2522, Australia

$^3$School of Physics \& Astronomy, University of Birmingham, Birmingham B15 2TT, UK

$^4$Massachusetts Institute of Technology, Center for Space Research, Cambridge, MA 02139

$^5$Dept. Physics, University of Bristol, Tyndall Avenue, Bristol BS8 1TL, UK

$^6$Institute for Astronomy, 2680 Woodlawn Drive, Honolulu, HI 96822

$^7$University of Manchester, Jodrell Bank Observatory, Macclesfield, Cheshire SK11 9DL, UK

$^8$Dept. Physics and Astronomy, Johns Hopkins University 3400 N. Charles Street, Baltimore, MD 21218

$^9$http://asc.harvard.edu/udocs/docs/docs.html

\normalsize

\noindent
power law spectra we follow
the convention that the flux density is S=$\rm{k}\nu^{-\alpha}$.

\section{Data Analysis}

The calibration observation of 3C295 was performed on 1999 Aug30 for
an elapsed time of 20,408s.  The target was near the aim point on the
S3 ACIS chip. We generated a clean data set by selecting the standard
grade set (0,2,3,4,6), energies less than 10~keV, and excising times
with enhanced background rates.  The screened exposure time for the
observation is 17,792~s.

During standard processing, photon events are assigned
fractional pixel locations due to spacecraft dither, rotation between 
detector and sky coordinates, and an additional
randomization within each $0''\llap{.}5$ ACIS pixel.
We used the fractional pixel values to
generate images with $0''\llap{.}1$ pixels.
Figure 1 shows the central region
after adaptive smoothing with a Gaussian constrained to have
$\sigma\geq0''\llap{.}3$.  Overlaid are the radio contours from a 20
cm MERLIN+VLA image (Leahy et al., in preparation).

\subsection{X-ray Morphology}

The ACIS-S image in Figure 1 shows that the central region of the
3C295 cluster exhibits significant structure with an X-ray core and
two outer features aligned with the radio hotspots.  This observation
is an excellent example of the resolving power of Chandra.  3C295 was
previously observed by the Einstein HRI (Henry \& Henriksen 1986) and
the ROSAT HRI (Neumann 1999) but the spatial resolution of these
instruments was comparable to the separation of the two X-ray hotspots
and these features were thus not detected.  The ACIS image in Figure 2
shows that the core and two hotspots are not simple, azimuthally
symmetric features, suggesting that these regions are resolved by
Chandra.  To determine if the core and hotspots are extended we must
model the cluster emission. The region outside a $3^{\prime\prime}$
radius is well fit with a $\beta$ model with $\beta$=0.54 and core
radius $a= 4''\llap{.}1$ (24~kpc).  The residual X-ray data above this
model are shown superimposed on the HST image in Figure~2. Inside
$r=3^{\prime\prime}$ there is clearly excess emission above the
$\beta$ model, surrounding the nuclear ~ and ~ hotspot ~ sources. ~ We

\begin{center}

\psfig{file=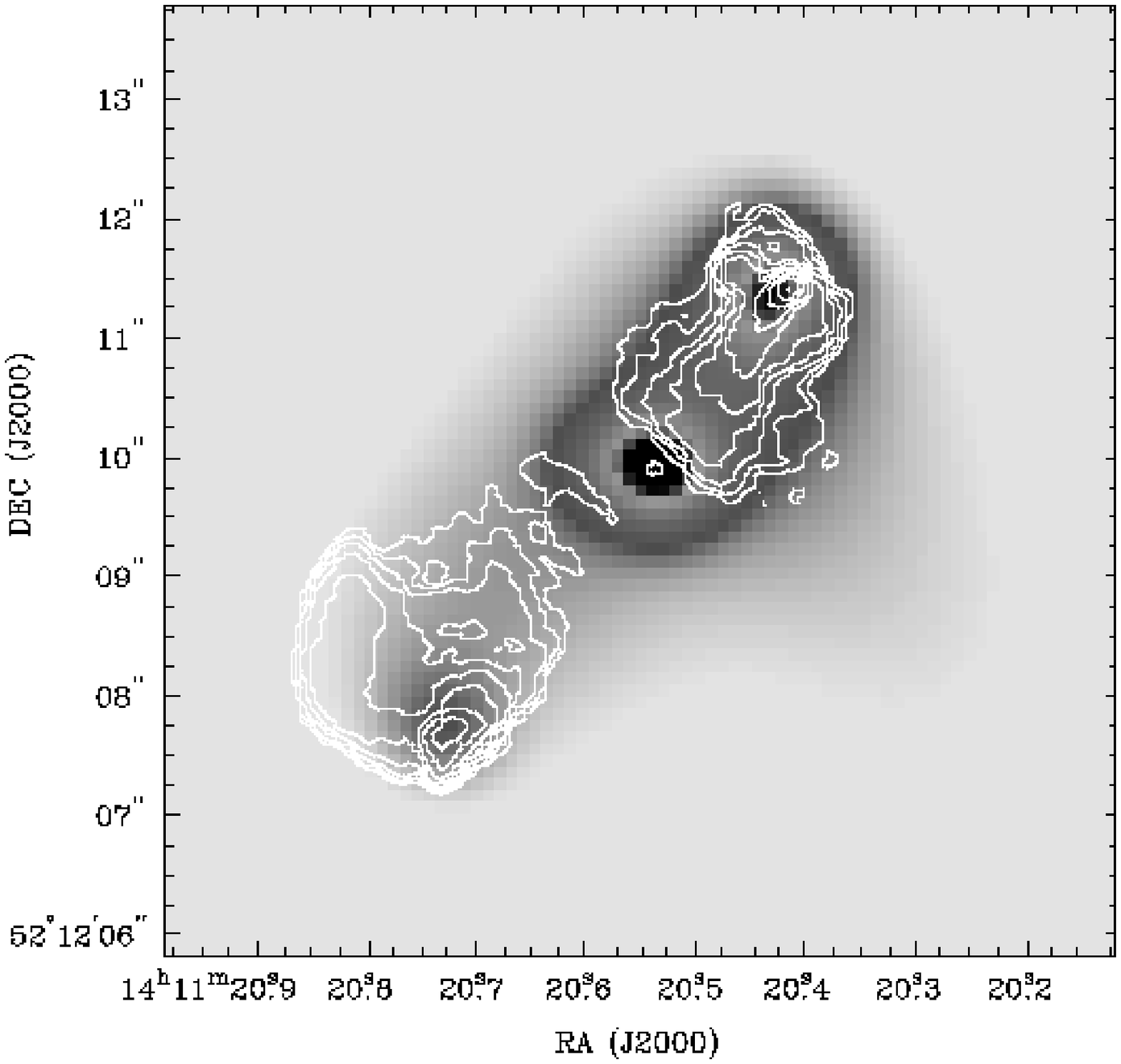,height=7.0cm,width=7.0cm}
\begin{minipage}{8cm}
\small\parindent=3.5mm {\sc Fig.}~1. The 20 cm MERLIN data (contours)
overlaid on the ACIS image.  The radio contours are logarithmically
spaced from 4 mJy to 2.0 Jy, shown with a restoring beamsize of
$0''\llap{.}14$.  The X-ray data have been shifted by
$0''\llap{.}66$ to align the X-ray and radio cores.  The NW hotspot 
is at a projected distance of $1''\llap{.}9$ (11 kpc) from the
core and the SE hotspot is $2''\llap{.}75$ (16 kpc) from the core.

\end{minipage}

\end{center}

\smallskip

\noindent
attempted to remove this by fitting a double $\beta$ model to the data
with the nuclear and hotspot regions excluded.  This removed the
surrounding emission fairly effectively, allowing us to study the size
of the central components.  To test these components for intrinsic
extension, we compared the ratio of net counts in two concentric
regions centered on these features in the residual image.  The ratios
of net counts between 0 and $0''\llap{.}5$ and from 0.5 to
$1''\llap{.}0$ for the northwest (NW) hotspot, and southeast (SE)
hotspot are $1.35\pm0.38$ and $0.97\pm0.35$, respectively (Poisson
errors).  We use the ACIS-S image of PKS0637-72 to determine this
ratio for an imaged point source, finding a ratio of $1.93\pm0.08$.
Based on this statistic, there is tentative evidence
in the Chandra data for some extension of the hot spot X-ray sources.
However, this result is subject to uncertainties in the shape of
the underlying diffuse distribution. In the case of the nuclear
source these systematics are dominant, and no useful extension test
is possible.

\subsection{X-ray Spectral Results}

The response matrices for the S3 chip have been calibrated on 32 by 32
pixel regions.  There are also twelve effective area files covering
the S3 chip.  A separate photon weighted response 

\small

\begin{center}

TABLE 1

Summary of Spectral Analysis Results

\begin{tabular}{lllll}
&Cluster & Nucleus & NW & SE \\
\hline\hline
Region &$r$=90$^{\prime\prime}$ &$r=0^{\prime\prime}\llap{.}9$
&$1^{\prime\prime}\llap{.}5 \times 1^{\prime\prime}\llap{.}5$ &
$r$=$1^{\prime\prime}$ \\

Net counts&4856$\pm$104& 137$\pm$15 & 138$\pm$14 &  42$\pm$13 \\

Model&Thermal&PL & PL & ... \\

kT or $\alpha$&$4.4\pm$0.6&-0.8$\pm$0.3 &0.9$\pm$0.5 & ... \\
$\chi^{2}$/DOF& 173/143& 5.3/4 & 3.1/3 & ... \\

Flux & 120. &19.0 &3.8 &1.1 \\

Lum& $9.6\times 10^{44}$&$7.3\times 10^{43}$ & $2.9\times
10^{43}$ &$9\times10^{42}$\\
\hline
\end{tabular}

\end{center}

Notes: The temperatures are given in keV; the confidence ranges for
the spectral parameters are for $\Delta\chi^{2}$=2.7 (90\% for one
parameter).  The fluxes are 'unabsorbed' for the 0.2 to 10 keV band at
the Earth in units of $10^{-14}$~erg~cm$^{-2}$s$^{-1}$.  The source's
rest frame luminosity in the 0.2 to 10 keV band is given in ergs
s$^{-1}$.

\normalsize 

\begin{center}

\psfig{file=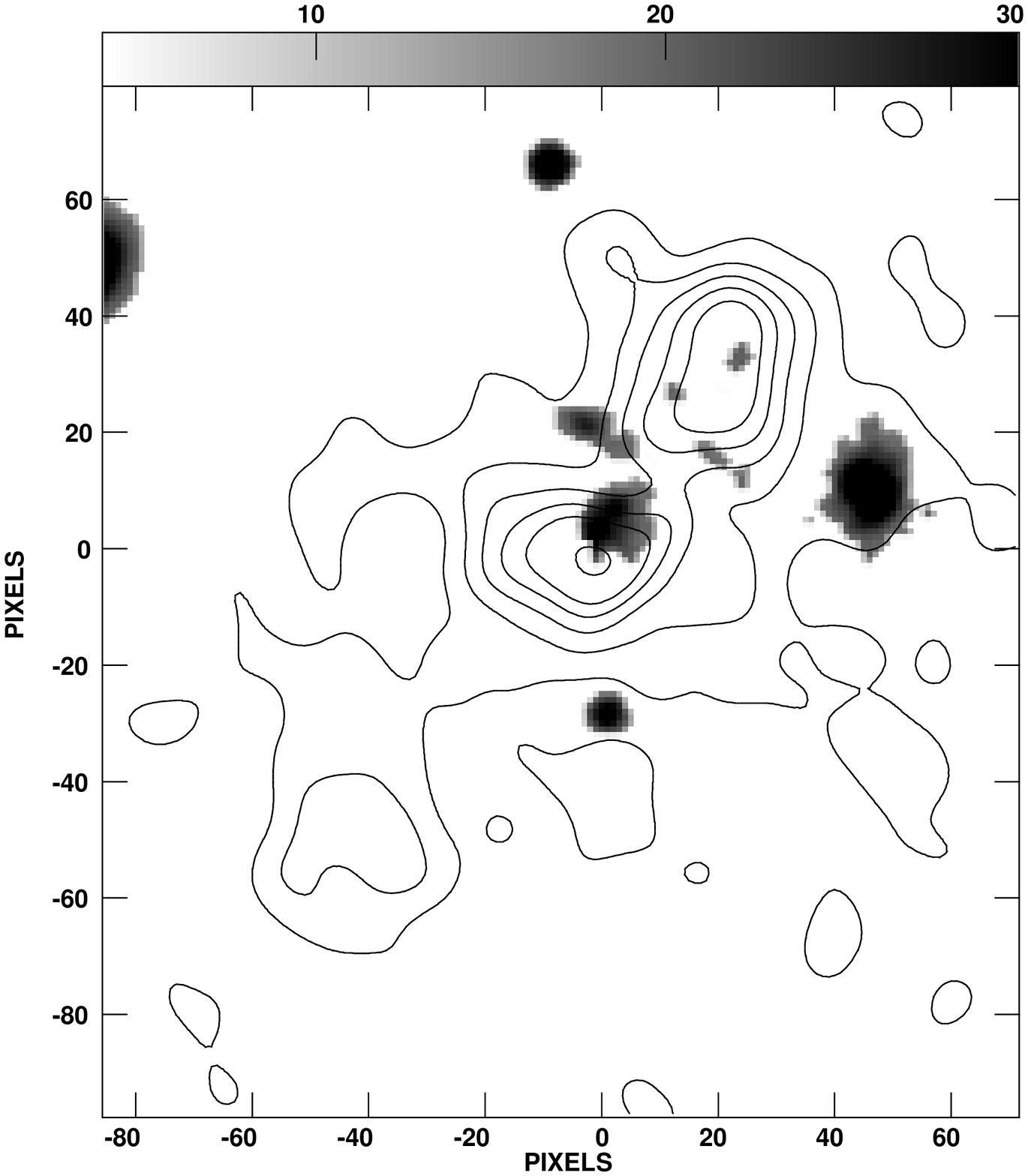,height=7.0cm,width=7.0cm}
\begin{minipage}{8cm}
\small\parindent=3.5mm {\sc Fig.}~2. A contour map of the residual Chandra image after subtraction
of the best fit $\beta$ model superimposed on the residual optical
emission from an HST observation after subtracting the emission
from the central galaxy.  The X-ray data
have been smoothed with a Gaussian function of FWHM=$0''\llap{.}5$.  Contour
levels are 8, 24, 40, 56, 72, 104 counts per square arcsecond. Coordinate
pixels are $0''\llap{.}045$.

\end{minipage}

\end{center}

\noindent
matrix and area file was generated for each extracted spectrum 
based on the chip coordinates of the detected photons. Spectral 
bins were chosen to
include at least 25 net counts per bin, and the data were fit over the
energy range from 0.5 to 7 keV.  The results are summarized in Table 1.

\subsubsection{The cluster gas}

A spectrum was extracted for the whole cluster within
$90^{\prime\prime}$ of the radio nucleus, excluding the central radio
source, hotspots, and two background sources.  A background spectrum
was extracted $\sim5'\llap{.}3$ northwest of the cluster center.  An
absorbed single temperature thermal model, with $N_{\rm H}$ fixed at
$1.34\times10^{20} \rm\,cm^{-2}$ (the galactic value) provides an
adequate fit for abundances from 0.2 to 0.5 cosmic.  The results in
Table 1 are based on a fixed value of 0.3 for the abundance.  Our
temperature is not formally consistent with the $7.13^{+2.06}_{-1.35}$~keV
obtained from analysis of ASCA data by Mushotzky \& Scharf (1997) but
our analysis is confined to the central $90^{\prime\prime}$
and excludes the emission from the core and two hotspots.  A more
detailed discussion of the cluster emission will be presented in a
subsequent paper.

\subsubsection{The hotspots}

There are insufficient counts to perform a spectral analysis of the SE
hotspot.  For the NW hotspot, a spectrum was extracted from a
$1''\llap{.}5$ square around the X-ray peak, centered $1''\llap{.}9$
from the nucleus.  There are no counts above 3.5 keV and only 5 bins
are used in the fit.

In order to minimize contamination by cluster thermal emission we used
a background spectrum extracted from an identical area at a position
$1''\llap{.}8$ SW from the nucleus, in a direction perpendicular to
the radio jets.  The results for a power law fit are given in Table 1.
A thermal model (with $N_{\rm H} =1.34\times10^{20} \rm\,cm^{-2}$ and
$Z = 0.3$ fixed) gives a best fit temperature of $kT = 4.4{+19 \atop
-2.2}$ keV and $\chi^2 = 5.1$ for 3 degrees of freedom, which is also
acceptable.

\small

\begin{center}

TABLE 2

Uniform Density Hot Gas:  Thermal Model Parameters

\begin{tabular}{lllcccc}
&Size   &  Mass  &      $n_e$    &P      &RM \\
&(arcsec)&(M$_{\odot}$)&(cm$^{-3}$)&(erg~cm$^{-3}$)&(rad/m$^2$) \\
\hline\hline
NW &$r$=0.1&$2.9\times10^8$&12&$1.7\times10^{-7}$&58,000. \\
&$r$=0.75&$5.9\times10^9$&0.60&$8.2\times10^{-9}$&21,000. \\
&&&&&& \\
SE&$r$=0.1&$1.6\times10^8$&6.8&$9.2\times10^{-8}$&32,000. \\
&$r$=0.75&$3.3\times10^9$&0.33&$4.5\times10^{-9}$&12,000. \\
\hline
\end{tabular}

\end{center}
Notes: the total mass (Mass), electron number density ($n_e$),
and pressure (P) required to reproduce the observed X-ray emission,
assuming a temperature of 4.4 keV.  The computed rotation measure
(RM) assumes B=10$\mu$G and a path length equal to $r$.

\normalsize

\smallskip

\subsubsection{The nucleus}

An X-ray spectrum was extracted from a region
$0^{\prime\prime}\llap{.}9$ in radius around the central peak.  In
order to remove cluster thermal emission, the background for this
spectrum was taken to be the same as that for the northern hotspot.  The
spectrum is extremely hard; most of the photons have E$>$1keV and
about half have E$>$3keV.

An absorbed power law fit provides an adequate representation with the
best fitting column density consistent with that from galactic
foreground absorption.  Thermal fits are unacceptable with
$\chi^{2}$ per degree of freedom $>34/3$.  Since the best fit value for $\alpha$
corresponds to a rather steep inverted spectrum, we also attempted
fits with column densities up to 10$^{24}$ cm$^{-2}$, but these all
gave a worse fit due to the low energy photons in the spectrum.  This
could be caused by residual thermal emission, but there are not enough
events to pursue this further.

\subsection{Other data}

MERLIN and VLA data between 1.5 and 43 GHz were taken from Perley \&
Taylor (1991) and Leahy et al. (in preparation).

We obtained archival HST data and after subtracting the emission from
the central galaxy obtained a flux of S=0.078 $\mu$Jy at
$\nu$=4.32$\times$10$^{14}$ Hz for the NW hotspot (with a one $\sigma$
uncertainty of 20\%) within a $0''\llap{.}1$ radius aperture. The
optical emission from the hotspot appears to be extended on the same
scale as the brightest radio structure and there are fainter features
extending back about $0''\llap{.}5$ towards the galaxy center.  The SE
hotspot is marginally detected at $3 \sigma$ with a flux density of
S=0.02$\mu$Jy (not visible in Figure 2).

\section{Emission processes for the hotspots}

\subsection{Thermal model for the NW hotspot X-ray emission}

We have calculated the electron density required to produce the
observed $L_x$ for a temperature of 4.4 keV and two representative
volumes: (a) for a 
sphere with radius $0''\llap{.}1$ (i.e. the 'unresolved'
case) and (b) a sphere with radius $0''\llap{.}75$
(the maximum size allowed by the data).  The
resulting values are shown in Table 2, along with the
excess rotation measure predicted for a
B field component along the line of sight of 10 $\mu$G and a
path length equal to the radius of the sphere.

Combining the best fit $\beta$ model and the X-ray luminosity of 3C295
yields a central electron density of $0.083 \rm\, cm^{-3}$.  For a
temperature of 4.4 keV, this gives an ambient gas pressure of
$1.1\times10^{-9} {\rm\, erg\, cm^{-3}}$.  The minimum pressure in the
NW hotspot is $7\times10^{-8} {\rm\, erg\, cm^{-3}}$ (Taylor \& Perley
1992), and hence the jet could drive a shock into the surrounding ~ gas.
~ Applying ~ the

\begin{center}

\psfig{file=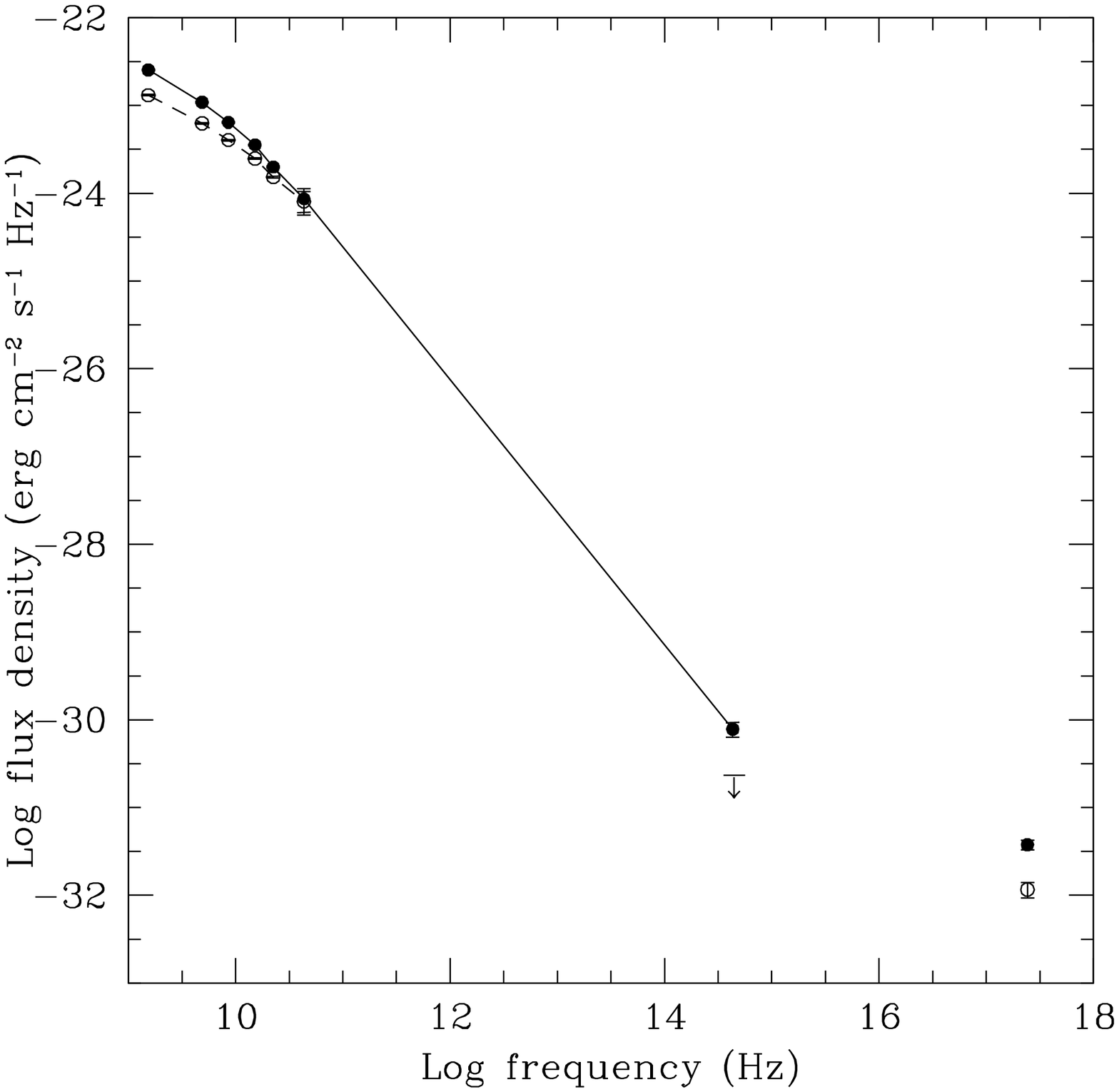,height=7.0cm,width=7.0cm}
\begin{minipage}{8cm}
\small\parindent=3.5mm {\sc Fig.}~3.  The observed spectrum of the 
hotspots.  The data for the NW and SE hotspots are shown as filled 
and open circles, respectively. The $3\sigma$ optical flux density for
the SE hotspot is shown as an upper-limit.  The X-ray data are plotted at 1
keV (log$\nu$=17.38).
\end{minipage}

\end{center}

\normalsize

\smallskip

\noindent
shock jump conditions for a postshock pressure of
$8.2\times10^{-9}\rm\, erg\, cm^{-3}$ (the most favorable case from
Table 2), gives a postshock density and temperature of $0.22\rm\,
cm^{-3}$ and 12 keV, respectively. 
Although this high temperature is
allowable within our 90\% confidence range, the density is still below
the values required to account for the X-ray luminosity (Table 2), and
the cooling time of the shocked gas ($\sim 3\times10^8$ yr) is far too
long for post-shock cooling to allow a rise in density.  For a smaller
X-ray emitting region the problem becomes more acute, and we conclude
that it is unlikely that the X-ray emission is due to shocked hot gas.

In addition, the largest allowed change in rotation measure (RM)
between the NW hotspot and its surroundings is about 2000 rad m$^{-2}$
(Perley \& Taylor 1991).  At the redshift of 3C295, this converts to
an intrinsic RM$\approx$4000 rad m$^{-2}$.  A somewhat smaller excess
is allowed for the SE hotspot.  These values are significantly less
than the predicted RMs in Table 2.  While multiple field reversals
along the line of sight to the hotspots could reduce the predicted
values, we note that the observed RMs are fairly constant over spatial
scales of 2 kpc, particularly for the NW hotspot. There are thus
several problems with a thermal origin for the emission from the NW
hotspot.

\subsection{Synchrotron Models}

Successful synchrotron models have been presented for knot A in the
M87 jet (Biretta, Stern, \& Harris 1991) and hotspot B in the northern
jet of 3C 390.3 (Harris, Leighly, \& Leahy 1998).  
An extension of the power laws from lower frequencies requires that
the electron population responsible for the radio (and optical)
emission extends to a Lorentz factor $\gamma$=10$^{7}$.  Extrapolating
from the radio/optical spectrum under-predicts the observed X-ray
emission by a factor of 500 for the NW hotspot, so a simple synchrotron
model is unacceptable.  For the SE hotspot, the discrepancy is more than a
factor of 1000 (see Figure 3).

Although we have not calculated synchrotron spectra from proton
induced cascades (PIC; Mannheim, Krulis, \& Biermann 1991), we suspect
that such a model would be feasible.  The primary difference between
the PIC and the SSC models is that PIC involves a very high energy
density in relativistic protons 

\small

\begin{center}
TABLE 3

Synchrotron Input Spectra for SSC Calculation

\begin{tabular}{lcccccc}
component&$\nu_1$&$\nu_b$&$\nu_2$&$\alpha_l$&$\alpha_h$&S$_b$ \\
&(Hz)&(Hz)&(Hz)&&&(cgs) \\ 
\hline\hline
NW&1E9&1E10&1E15&0.70&1.50&7.08E-24 \\
SE&1E9&1.33E10&1E15&0.70&1.58&3.16E-24 \\
\hline
\end{tabular}

\end{center}
Notes: $\nu_b$ is the frequency at which the spectral slope changes
and S$_b$ is the flux density at $\nu_b$. The radio spectrum is the
peak brightness as observed with a beam size of $0''\llap{.}2$ (large
enough to include the brightest structure seen at 43 GHz, but not so
large as to include a lot of surrounding emission).  The spectrum is
extended to 10$^{15}$ Hz to accommodate the HST data although this has
little effect on the derived parameters for the SSC model.  The radio
spectra are strongly curved (Fig. 3), which cannot be explained either
by self-absorption or entirely by spectral aging (given the optical
emission). So, following Carilli et al. (1991), we assume that the
electron energy spectrum cuts off near the bottom of the observed
band. Flux densities are given in ergs~cm$^{-2}$~s$^{-1}$~Hz$^{-1}$.

\normalsize

\bigskip

\noindent
which would indicate a much higher B
field (i.e. $>1000\mu$G) than those estimated from the minimum energy
conditions assuming the electrons are the major contributor to the
particle energy density.

\subsection{Synchrotron Self-Compton Model}

The radio structure of the hotspots
is quite complex, and if we estimate photon energy densities for the
brightest regions (from the 43GHz VLA data), we would expect the
ACIS detections to be unresolved.  Since it is difficult to verify
this, we base our calculations on the small volumes, realizing that
there may be additional, weaker contributions from somewhat larger
scale features (weaker because the photon energy density will be
smaller).

Our estimates involve defining the radio spectrum for the brightest
and smallest structure of the hotspots (Table 3) and then calculating
the synchrotron parameters: the minimum pressure magnetic field,
B$_{minP}$, the luminosity, L$_{sync}$, and the photon energy density,
$u(\nu)$.  The spectral coverage of SSC will be determined by
$\nu_{out}\approx\gamma^2\times\nu_{in}$, and the ratio of energy
losses in the IC and synchrotron channels will be
$R~=~u(\nu)/u(B)~\approx~L_{ic}~/~L_{sync}$, where $u(B)$ is the
energy density in the magnetic field.  We can then compute the
amplitude of the IC spectrum for which $\alpha$(x)= $\alpha$(radio)
and which contains the proper luminosity over the defined frequency
band.

The calculated SSC values (Table 4) demonstrate that the
results are not particularly sensitive to the assumed geometry and
that the SSC process is most likely the major contributor to the
observed X-ray intensities.  In addition to the values shown in
Table 4, an SSC estimate for the more extended radio structures of the
NW hotspot (i.e. extending $\approx0''\llap{.}4$ back towards the nucleus)
provides an additional 10\% of the total observed flux density at
1 keV.

For magnetic field values of $410 \mu G$ and $310 \mu G$ in the NW and
SE hotspots (slightly less than the equipartition values in Table 4),
the SSC model can account for the entire observed X-ray emission.  For
even lower magnetic fields, more relativistic electrons would be
required, which would result in a greater X-ray luminosity than
observed.  Hence, these magnetic field values are strict lower limits.
The near coincidence of the magnetic field value required by the SSC
model and equipartition 
lends weight to the assumption of
equipartition and suggests that the proton energy density is
not dominant. 

\small

\begin{center}
TABLE 4

SSC parameters for the hotspots

\begin{tabular}{llccccc} 
&geom.&B$_{minP}$&u($\nu$)&R&S(1keV)&S(ssc)/S(obs) \\
&&($\mu$G)&(erg cm$^{-3}$)&&(cgs) & \\ 
\hline\hline

NW&sphere&319&4.4E-10&0.11&2.1E-32&0.57 \\
NW&cylin.&561&1.6E-9&0.13&1.9E-32&0.51 \\
&&&&&& \\
SE&sphere&271&2.3E-10&0.08&1.3E-32 &1.16 \\
SE&cylin.&336&2.9E-10&0.06&9.7E-33&0.84 \\
\hline
\end{tabular}

\end{center}
Notes: For the sphere, $r=0''\llap{.}1$ (to match the radio
beamsize).  The cylinder for the NW hotspot has $r=0''\llap{.}043$ and
$l=0''\llap{.}1$.  For the SE hotspot, $r=0''\llap{.}05$ and
$l=0''\llap{.}25$.  B$_{minP}$ is the minimum pressure field for no
protons and filling factor=1.

\normalsize

\medskip

\section{Conclusions}

The SSC model provides good agreement between the classical
equipartition magnetic field estimates and the average field values
required to produce the observed X-ray intensities.  Since the
equipartition estimates were made with no contribution to the particle
energy density from protons, this agreement supports 
the hypothesis that relativistic electrons account for a
major part of the the energy density.  If
future observations show that the X-ray emission is actually extended
over a physical distance of 9 kpc, it is unlikely 
that SSC can be the sole process operating,
because the photon energy density will be much lower outside
the compact core of the radio hotspot.

SSC is a mandatory process and the only uncertainty in the predictions
is the value of the magnetic field.  As was the case for the hotspots
of Cygnus A, the only viable model to negate these conclusions is one
with a much stronger magnetic field: i.e. significantly greater than
500 $\mu$G.

\section*{Acknowledgments}

A complex observatory such as Chandra represents a tremendous effort by an
extensive team.  We thank in particular the ACIS PI, G. Garmire,
who was responsible for the detector which allowed us
to obtain our results.  
The work at CfA was partially supported by NASA contracts NAS5-30934
and NAS8-39073. PM gratefully acknowledges support from CNAA bando
4/98.  PEJN and TJP gratefully acknowledge the hospitality of the
Harvard-Smithsonian Center for Astrophysics.

\section*{References}

\ref{Biretta, J.A., Stern, C.P., \& Harris, D.E. 1991, AJ., 101, 1632}

\ref{Carilli, C.L., Perley, R.A., Dreher, J.W., \& Leahy, J.P. 1991,
\apj, 383, 554}

\ref{Harris, D.E., Leighly, K.M., \& Leahy, J.P. 1998, ApJ., 499, L149}

\ref{Harris, D.E., Carilli, C.L., \& Perley, R.A. 1994, Nature, 367, 713} 

\ref{Henry, J. P. \& Henriksen, M. J.  1986, ApJ, 301, 689}

\ref{Mannheim, K., Krulis, W.M., \& Biermann, P.L. 1991, \aap, 251, 723}

\ref{Mushotzky, R.F. \& Scharf, C.A. 1997, ApJ., 482, L13}

\ref{Neumann, D. M. 1999, ApJ., 520, 87}

\ref{Perley, R.A. \& Taylor, G.B. 1991, \aj, 101, 1623}

\ref{Taylor, G.B. \& Perley, R.A. 1992, \aap, 262, 417}

\vskip 0.9in

\end{document}